\documentclass{PoS}

\title{Iron line emission from X-ray pulsars: physical conditions and geometry of the system}

\ShortTitle{Iron line emission from X-ray Pulsars}

\author{\speaker{Sergey Tsygankov}\\%
        MPI for Astrophysics, Munich, Germany;\\
        Space Research Institute, Moscow, Russia\\
        E-mail: \email{sst@mpa-garching.mpg.de}}

\author{Alexander Lutovinov\\
        Space Research Institute, Moscow, Russia\\
        E-mail: \email{aal@iki.rssi.ru}}

\abstract{

We present here the preliminary results of the study of the fluorescent iron
line emission from X-ray pulsars with Be companions. We propose to use
properties of this emission to investigate the spatial distribution and
physical conditions of the matter around the compact object as well as in
the binary system as a whole.  Using data of the RXTE observatory the iron
line behavior in the transient X-ray pulsar V\,0332+53 spectrum was studied
during the powerful type II outburst in 2004$-$2005.  Particularly, we
investigated a variability of the iron line equivalent width on different
time scales (pulse period, orbital period, outburst phase) and searched for
its correlation with the continuum flux, spectral parameters, etc. }

\FullConference{The Extreme sky: Sampling the Universe above 10 keV\\
		 October 13-17 2009\\
		 Otranto (Lecce) Italy}

\begin{document}

\section{Introduction}

The fluorescent 6.4 keV iron line is broadly detected in spectra of
different types of astrophysical objects. Its properties were described
theoretically for different geometries of an emitting/reprocessing system
(e.g. \cite{bas80},\cite{gf91},\cite{lc93} and references therein). And it
was shown that this spectral feature can be a powerful instrument for the
study of the spatial distribution and state of the matter around X-ray
sources.

The distribution of matter and its state are crucial components for the
understanding of the mechanism of giant (type II) outbursts, observed from
X-ray transient pulsars with Be optical companions. Type II X-ray
outbursts ($L_x>10^{37}$ erg s$^{-1}$) last several weeks or even
months and are not correlated with any particular orbital phase
(\cite{neg98} and references therein).
Despite of many years of
observations the exact mechanism of such events 
as well as the origin and distribution of the
matter in the system to be accreted onto the compact object in each
particular moment are still unclear. On the other side a very promising
property of such sources is a wide range of observed luminosities and hence
the possibility to investigate a response of the system (particularly,
changes in the geometry) on different accretion rates. It is important to
note that such transient systems are usually too bright to be investigated
with grazing-incidence X-ray telescopes. But, fortunately, a large amount of
high quality data collected up to day from the RXTE observatory gives us a
possibility to trace the evolution of the fluorescent iron line emission on
different time scales (pulse period, orbital period, outburst phase) in
Be-NS transient systems and to try to answer to above mentioned questions.

In this report we present preliminary results of such a study of the very
bright transient X-ray pulsar V\,0332+53 during the 2004-2005
outburst. V\,0332+53 is a classical transient X-ray pulsar with Be
companion. The main system parameters are as follows: the
pulse period $\sim4.375$ s, the orbital period -- $34.67$ days, the
eccentricity -- 0.37, and the projected semimajor axis of the neutron
star -- $a_{x}$sin$i\simeq86$ lt-s (\cite{st85,zh05}).
Detailed analysis of this and other bright transient pulsars will
be published in a forthcoming paper (Tsygankov et al., in preparation).

\section{Observations and data analysis}

In the following analysis we used data from the PCA spectrometer onboard the
RXTE observatory \cite{br93} (a list of pointings IDs can be found in our
previous papers \cite{ts06},\cite{ts10}). The PCA spectrometer is a system
of five proportional xenon/propane counters with an effective area of about
6400 cm$^{2}$ and an energy resolution of 18\% at 6--7 keV. Despite of the
moderate energy resolution this instrument has a great advantage due to its
large effective area and high temporal resolution, which are of great importance
for the investigation of the variability of continuum and iron line fluxes
on short time scales (see below). Moreover, it gives us a possibility to
apply the potential of the reverberation mapping method, proposed in AGNs
studies (\cite{rey99} and references therein), to investigate the
distribution of a cold matter around neutron stars. For the data reduction
we used standard programs of the FTOOLS/LHEASOFT 6.7 package.

\section{Results}

\begin{figure}
  \begin{minipage}{0.6\textwidth}
    \resizebox{\hsize}{!}{\includegraphics{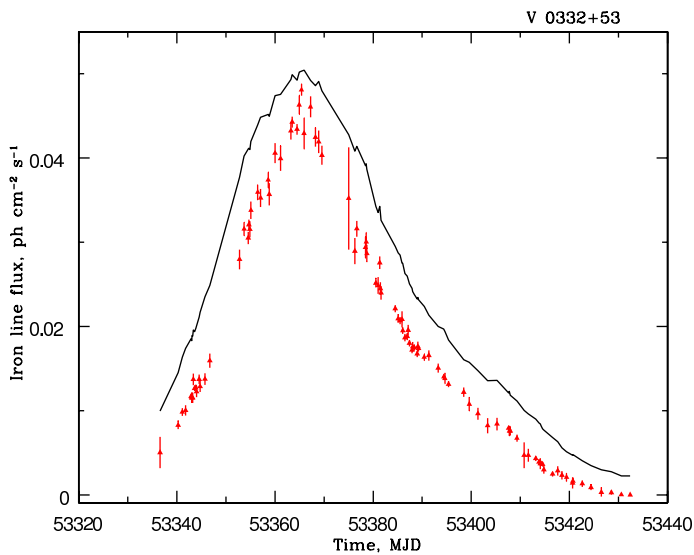}}
  \end{minipage}
  \begin{minipage}{0.35\textwidth}
    \caption{
Dependence of the iron $K_{\alpha}$ line (points) and continuum
fluxes in the 4-12 keV energy range (solid line, arbitrary units) on the
time during the outburst from V\,0332+53 in 2004-2005.}\label{fenorm}
  \end{minipage}
\end{figure}


\subsection{Long-term variability}

In Fig.\ref{fenorm} the evolution of the iron line flux and the continuum
one in the energy range of $4-12$ keV are presented. It is seen that
observed variations of the iron line flux are mainly caused by variations of
the continuum flux. An analysis of a behavior of the iron line equivalent
width (EqW) can be significantly more informative because this quantity bear
the impress of geometrical properties of the system  (e.g.
\cite{bas80},\cite{gf91},\cite{lc93}) and doesn't depends on
the incoming flux.

As it was shown in \cite{ts06},\cite{ts10} the source spectrum in a wide
energy band has a complex shape and has been described with a large number
of parameters. Therefore to avoid a possible influence of other spectral
parameters on the iron line ones we restricted here our consideration by a
relatively narrow energy band (3-12 keV) of the PCA spectrometer, where the
continuum emission can be well approximated by a simple power law model with
an exponential cutoff. To describe the iron line emission we added to our
model a Gaussian component. The line center and its normalization were
left free for the fitting, but its width was fixed at
$0.2$~keV. Nevertheless, the absolute value of the equivalent width,
obtained from the spectral analysis, was still model dependent, that is
connected with a moderate energy resolution and relatively high lower energy
limit of the spectrometer ($\sim$3 keV). Therefore below we will
concentrated our attention to relative changes of EqW.

Note, that a variability of the intensity and equivalent width of the iron
line was observed before for several X-ray pulsars on the long time scales
(of the order of the orbital period). But these variations were caused
mainly by changes of the density of the strong stellar wind around the
compact object.  In particular, Inoue \cite{in85} showed that the expected
equivalent width of the iron K$_{\alpha}$ line is a linear function of the
hydrogen column density and could be expressed as (for a spherical geometry
and photon index of 1.1):

\begin{equation}\label{eq_nh}
EqW=100\left(N_{H}/10^{23}\right) eV,
\end{equation}

where $N_H$ is in units of cm$^{-2}$.

\begin{figure}
  \begin{minipage}{0.6\textwidth}
    \resizebox{\hsize}{!}{\includegraphics{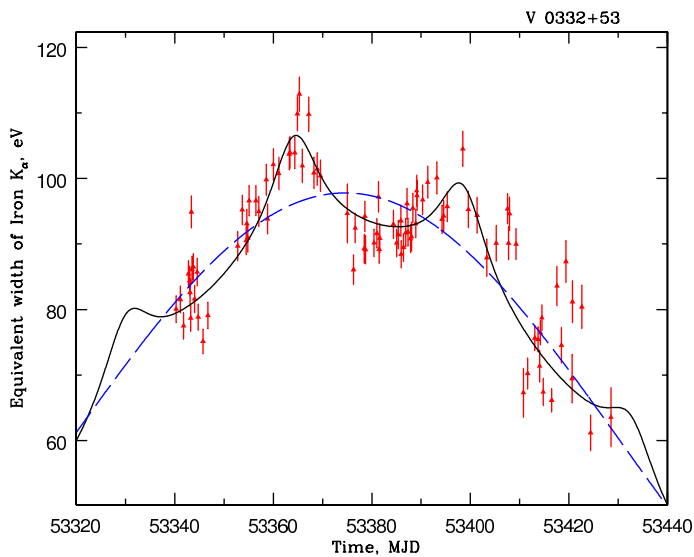}}
  \end{minipage}
  \begin{minipage}{0.35\textwidth}
    \caption{
Equivalent width of the iron $K_{\alpha}$ line versus time during
the outburst from V\,0332+53 in 2004-2005. Dashed line represent the
rough approximation of the luminosity dependence by the Gaussian law
(in arbitrary units), solid line -- its convolution with the inverse distance
between components of binary system.}\label{eqw0332}
  \end{minipage}
\end{figure}

The equivalent width of the iron line, registered in the V\,0332+53
spectrum, is shown as a function of time in Fig.\ref{eqw0332}. Its value was
changed from $\sim60$ to $\sim110$ eV during the outburst approximately in
the same way as the source luminosity, whose rough approximation by the
Gaussian law is presented by a dashed line in arbitrary units. If the
observed value of EqW is due to the reprocessing in the surrounding stellar
wind, then according to eq.\ref{eq_nh} a very high (about 10$^{23}$
cm$^{-2}$) value of the photoelectron absorption should be detected in the
source spectrum. In fact we did not see any evidences of it at any level of
the source luminosity.

One of the natural reprocessing site which is able to produce variations of
the iron line parameters over the orbital cycle is the companion star atmosphere
and circumstellar disk around it. Suggesting that the iron line equivalent
width is proportional to the angular size of this disk seen from the compact
object, we can estimate its geometrical size and trace its evolution with
outburst and orbital phases (note, that the last one reflects the distance
between components changing due to a significant eccentricity of the binary
system). In general, changes of the equivalent width during outburst can be
preliminary described by the convolution of the mentioned above Gaussian law with the
inverse distance between components (solid line in Fig.\ref{eqw0332}, T$_0$
value was chosen arbitrarily). It is seen that data points are described more
or less satisfactorily even by such a simple geometrical model.

It was shown by \cite{gf91} that EqW is strongly affected by the photon
index of the incident spectrum. During the outburst the source spectrum has
been harder in low luminosity states and this fact should be also taken into
account during the modelling.

\subsection{Variability on the pulse period time scale}

Pulsations in the iron K$_{\alpha}$ line were detected earlier from several
X-ray pulsars (Vela X-1 \cite{choi96}, Her X-1 \cite{zane04}, etc.). Such
pulsations naturally indicate that the distribution of the reprocessing
matter is not completely spherically symmetric around the neutron star and
some compact reprocessing regions in its vicinity should exist. By way of
such a reprocessor it can be proposed a surface of the neutron star,
accretion column, Alfv\'{e}n shell, accretion wake or accretion disk,
stellar wind and atmosphere of companion star, etc. In particular, measuring
the delay of the iron line emission from the continuum component in the
around energy band Kohmura et al. \cite{koh01} have concluded that the
reprocessing site in the X-ray pulsar Cen X-3 is the accreting matter along
the magnetic field around the neutron star at the distance of
$\sim2\times10^{8}$ cm.

\begin{figure}
\center{\includegraphics[width=0.9\columnwidth, clip]{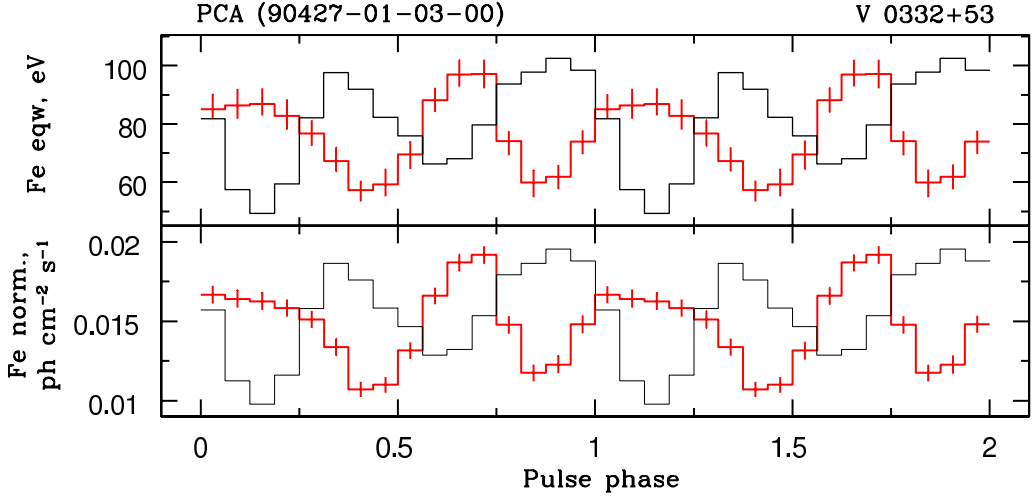}}

\caption{Dependence of the equivalent width (top panel) and normalization
(bottom panel) of the iron $K_{\alpha}$ line in the V\,0332+53 spectrum on
the pulse phase during the observation of ID 90427-01-03-00.}\label{phres}

\end{figure}

We propose to apply the pulse-phase resolved spectroscopy and cross-correlation
analysis methods to try to measure such a delay for V\,0332+53 and answer to
the question about a distribution of the matter in the system. A detailed
description of the phase-resolved spectroscopy results over the whole
outburst will be published in a separate paper (Lutovinov, Tsygankov,
prepared for publication). Here we present only the evolution of iron 6.4
keV line parameters with the pulse phase.

In Fig.\ref{phres} the iron line equivalent width and normalization changes
(red histograms) are presented in line with ones of the continuum flux
(black histograms) over the pulse period during the RXTE observation
90427-01-03-00, when the source luminosity was
$\sim\!2\!\times\!10^{38}$~erg s$^{-1}$. It is clearly seen that both EqW
and normalization are shifted relative to the pulse profile on about of 1.5
sec. Considering the measured time lag as a time-of-flight of the continuum
photons from the source to the reprocessing site it is possible to estimate
a distance between them as $\sim5\times10^{10}$ cm. This distance is about 2
orders of magnitude higher than the Alfv\'{e}n surface radius and comparable
with the outer size of the accretion disk.

Another possible explanation of the observed time lag is a phase lag
(arising just due to a rotation of the neutron star). In this case the
illuminating beam cuts the reprocessing site later than it cuts a direct
line of the sight to the observer.  Mentioned above cross-correlation
analysis will help us to determine a real value of the distance between
X-ray source and fluorescent matter.

Finally note that the pulsed fraction of the iron line emission during this
observation was about 30\%, that is significantly higher than its average
value in the 3-20 keV energy range \cite{ts10}.  Such a high value of the
iron emission pulsed fraction means that the reprocessing region size cannot
be too large. On the contrary, in the case of GX 301-2 the local decrease of
the pulsed emission at energies of 6-7 keV have been found by
\cite{endo02}. This fact can be easily explained by the reprocessing of the
iron emission in a thick 4$\pi$-surrounding stellar wind.

\section{Conclusions}

We found for the first time the correlation of the equivalent width of the
fluorescent iron line with the orbital phase during a huge type II outburst
from the transient X-ray pulsar V\,0332+53.

Pulsations in the iron line were found in a wide range of the source
luminosities, thus indicating a compact reprocessing site in the vicinity of
the neutron star.

From the pulse phase resolved analysis we measured a time lag of $\sim1.5$ sec
between continuum and iron line fluxes, that corresponds to the reprocessing
site at a distance of $\sim5\times10^{10}$ cm.

The significant increase of the pulsed fraction in the iron line emission in
a comparison with the continuum one was found for the first time.

\section{Acknowledgments}

Authors thanks M.Revnivtsev and D.Klochkov for useful discussions.  This
work was partially supported by the Program ``The origin, structure, and
evolution of objects of the Universe" of the Russian Academy of Sciences and
grant NSh-5069.2010.2 for support of leading scientific schools.  We are
grateful for the data to the HEASARC Online Service provided by the
NASA/Goddard Space Flight Center.

\end{document}